\documentclass[aps,prl,twocolumn]{revtex4}
\usepackage{graphicx,epsfig}
\def\be{\begin{eqnarray}}
\def\ee{\end{eqnarray}}
\def\vk{\vec k}

\begin{document}

\title{Self-energy analysis of frequency-dependent conductivity: \\
Application to Pb, Nb, and MgB$_2$ in normal state}
\author{Tae-Hyoung Gimm and
Han-Yong Choi\footnote{To whom the correspondences should be
addressed. e-mail: hychoi@skku.ac.kr}}

\affiliation{Department of Physics, BK21 Physics Research
Division, and Center for Nanotubes and Nanostructure Composites,
Sung Kyun Kwan University, Suwon 440-746, Korea.}

\date{\today}

\begin{abstract}

We propose and demonstrate a microscopic way to analyze the
frequency-dependent infrared conductivity: extraction of the
electron self-energy from the inversion of experimentally measured
infrared conductivity through the functional minimization and
numerical iterations. The self-energy contains the full
information on the coherent and incoherent parts of interacting
electrons and, therefore, can describe their charge dynamics even
when the quasi-particle concept is not valid. From the extracted
self-energy, other physical properties such as the Raman
intensity spectrum and the effective interaction between electrons
can also be computed. We will first demonstrate that the
self-energy analysis can be successfully implemented by fitting
the frequency-dependent condcutivities of the simple metals such
as Pb and Nb, and then calculating the effective interactions
between electrons from the extracted self-energies and comparing
them with those obtained from the tunneling experiments. We then
present the self-energy analysis of the MgB$_2$ superconductors
in normal state and clarify some of the controversies in their
optical spectra. In particular, the small electron-phonon coupling
constant obtained previously is attributed to an underestimate of
the plasma frequency.

\end{abstract}


\maketitle

\section{INTRODUCTION}

The frequency-dependent conductivity $\sigma(\omega)$ provides
one of the most valuable and detailed information on the charge
dynamics in a wide class of materials. It is analyzed using
either the one-component or two-component models \cite{tanner}.
The two-component model interprets the $\sigma(\omega)$ as arising
from a combination of two types of carriers, free and bound ones.
The free carriers are modeled in terms of the Drude term and the
bound ones in terms of various Lorentzian oscillators. However,
the interpretation of the individual Lorentzian terms, most of
which are usually due to inter-band contributions, is not
straightforward.

In the one-component picture, referred to as the extended Drude
model (EDM), on the other hand, the frequency dependence of the
conductivity $\sigma(\omega)$ below inter-band contributions is
described by extending the phenomenological parameters of the
Drude model, the effective mass $m^*$ and scattering rate
$1/\tau$, to be frequency dependent as \cite{tanner}
\begin{equation}
\label{drude} \sigma(\omega) = \frac{ne^2}{m_b}
\frac{1}{1/\tau(\omega)-i\omega m^*(\omega)/m_b} ,
\end{equation}
where $m_b$ is the electron band mass. The EDM interprets the
experimentally obtained complex conductivity $\sigma(\omega)$ in
terms of $1/\tau(\omega)$ and $m^* (\omega)$ determined by
 \be
 \label{EDMpara}
\frac{1}{\tau(\omega)}=\frac{\omega_p^2}{4\pi} Re \left[
\frac{1}{\sigma(\omega)} \right], ~~\frac{m^\ast(\omega)}{m_b}
=-\frac{\omega_p^2}{4\pi} \frac{1}{\omega}Im \left[
\frac{1}{\sigma(\omega)} \right],
 \ee
where $\omega_p = (4\pi n e^2/m_b)^{1/2}$ is the plasma
frequency, $n$ the electron density, and $e$ is the electron
charge. The $\omega_p$ can be found by integrating the real part
of the measured conductivity from the sum rule
 \be
\int_0^{\omega_{max}}d\omega \sigma_1(\omega) = \frac18
\omega_p^2,
 \label{sum-rule}
 \ee
where $\omega_{max}$ is the cutoff frequency above which
inter-band contributions begin to contribute. The subscript 1 and
2 refer to, respectively, the real and imaginary parts. The EDM
has been successfully employed to analyze $\sigma(\omega)$ of the
conventional metals as well as the heavy-fermions and high-$T_c$
cuprates \cite{puchkov}.

For a class of correlated electron systems such as the ruthunates,
Sr/CaRuO$_3$, however, the EDM breaks down and yields unphysical
descriptions of the materials such as the negative effective mass
\cite{kostic,yslee}. Similar behavior is also found for some
molybdates, Sm$_2$Mo$_2$O$_7$ and Nd$_2$Mo$_2$O$_7$ \cite{mwkim}.
These observations clearly signal the inadequacy of the EDM and
call for a new way of analyzing $\sigma(\omega)$ which can be
applied to a wide class of materials. We, therefore, propose to
analyze the frequency dependent infrared (IR) conductivity in
terms of the electron self-energy $\Sigma(\omega)$ instead of the
phenomenological parameters of Eq.\ (\ref{EDMpara}). The electron
self-energy contains the full information on the coherent and
incoherent parts of interacting electrons and, therefore, can
describe the charge dynamics even when the EDM or the Fermi
liquid (FL) picture is no longer valid. In this self-energy
analysis (SEA) method, the electron self-energy is extracted by
inverting the experimentally measured infrared conductivity
through the functional minimization and numerical iterations as
will be discussed in detail in Section III. The SEA may be
considered as a microscopic generalization of the EDM, which can
be applied to analyze the frequency-dependent conductivity data of
the non-Fermi liquids as well as the Fermi liquids. Even for the
FL where the EDM is expected to work, the SEA can yield
quantitatively more reliable results than the EDM, especially for
the FL with a strong electron-phonon coupling. The present paper
is mainly devoted to a detailed description of the SEA method and
its applications to relatively simple metals. For more complicated
cases of strongly correlated electron systems including cuprates,
molybdates and ruthunates, by which we were originally motivated,
we plan to report the SEA results separately elsewhere. For these
systems, qualitatively different results from the EDM results are
expected.

After the Introduction, we will discuss the frequency-dependent
conductivity $\sigma(\omega)$ expressed in terms of the
self-energy $\Sigma(\omega)$ and its relation with the EDM in
Sec.\ II. This will clarify the inherent limitations of the
widely employed EDM analysis for the frequency-dependent
conductivity. We will then describe in Sec.\ III the formulation
of the SEA method, which is reduced to the global minimization of
a $N$-variable function. The SEA method will be applied to
experimental data of Pb, Nb and MgB$_2$ and the results are
presented in comparison with the EDM analysis in Sec.\ IV. In
Sec.\ V, we give a brief summary and some perspectives on the
self-energy analysis method.

\section{Optical Conductivity and Self-energy}

The frequency-dependent conductivity $\sigma(\omega)$ can be
obtained from the current-current correlation function, and is
written in terms of the electron self-energy $\Sigma(\omega)$ as
\cite{nam,shulga}
 \be
\label{calculation} \sigma(\omega) =\frac{\omega_p^2}{4\pi}
\frac{i}{\omega}\int^\infty_{-\infty} d\epsilon \frac{f(\epsilon
- \omega) - f(\epsilon)}{\omega -
\Sigma(\omega-\epsilon)-\Sigma(\epsilon)} \/,
 \ee
where $f(\epsilon)$ is the Fermi distribution function. We assumed
a constant density of states over an infinite bandwidth and no
long-range order. It is also assumed that the momentum dependence
is much weaker compared with the frequency dependence,
$\Sigma(\vk,\omega) = \Sigma(\omega)$, as in the dynamical
mean-field theory, which renders the vertex correction vanish in
the current-current correlation function \cite{georges}.

The electron self-energy represents the effects of electron
interaction with various excitations in a system. The imaginary
part of the self-energy can be written as
\be
\label{a2F} \Sigma_2(\omega) &=& -\int^{\infty}_{-\infty}d\Omega
\left[\coth\left (\frac{\Omega}{2T} \right) \right . \nonumber \\
&& \left . + \tanh\left(\frac{\omega-\Omega}{2T}\right)\right] P_2(\Omega)
+ \Sigma_2^{imp}\/,
\ee
where $\Sigma_2^{imp}$ is the frequency-independent contribution
from impurities. $P_2(\omega)$ is the imaginary part of the
effective interaction satisfying $P_2(-\omega) = -P_2(\omega)$ and $T$ is the temperature. The
isotropically weighted phonon density of states for
electron-phonon coupled systems is given by
 \be
 \label{alpha2f}
\alpha^2 F(\omega) = \frac1\pi P_2 (\omega) .
 \ee
Once the self-energy is known, we use Eq.\ (\ref{a2F}) to find
the effective interaction spectrum $P_2 (\omega)$ using a
derivative with respect to $\omega$ at low $T$ or a convolution
after Fourier transformations \cite{shulga}.

In the zero temperature limit $T= 0$, the $\sigma(\omega)$ of Eq.\
(\ref{calculation}) is reduced to
 \be
 \label{sigmaT0}
\sigma(\omega) =\frac{\omega_p^2}{4\pi}
\frac{i}{\omega}\int_0^\omega d\epsilon \frac{1}{\omega -
\Sigma(\omega-\epsilon)-\Sigma(\epsilon)} \/,
 \ee
which means that in the low $T$ limit, only the self-energy
$\Sigma(\epsilon)$ between $0<\epsilon<\omega$ contributes to the
conductivity at $\omega$. This may be interpreted as
$\sigma(\omega)$ being an ``average'' of
$1/[\omega-\Sigma(\epsilon)-\Sigma(\omega-\epsilon)]$ between 0
and $\omega$. This, in turn, suggests that the information from
the EDM analysis, which is directly obtained from $\sigma(\omega)$
using Eq.\ (\ref{EDMpara}), is an average of the corresponding
quantity from the SEA. This will be discussed in more detail
below. The EDM can be obtained from Eq.\ (\ref{calculation}) in
an appropriate limit. The $\sigma(\omega)$ of Eq.\
(\ref{calculation}) is reduced to the EDM form of Eq.\
(\ref{drude}) with $m^*(\omega)/m_b = 1+\lambda(\omega)$ provided
that
 \be
 \label{EDMcondition}
\Sigma(\omega-\epsilon)+\Sigma(\epsilon) \approx
-\omega\lambda(\omega)-i/\tau(\omega)
 \ee
is satisfied \cite{shulga2}. For a FL, where $\Sigma_1(\omega)
\approx -\lambda(\omega) \omega$ and $\Sigma_2(\omega) \approx
\Sigma_2^{imp} -\Gamma \omega^2$, this condition can be satisfied
for small $\epsilon$ and $\omega $, if $\lambda(\omega)$ has a
weak $\omega$-dependence. Therefore, the EDM can give a
satisfactory description of $\sigma(\omega)$ for weak-coupling
FL, where
 \be
 \label{mapping}
\frac{1}{\tau(\omega)} \approx -2 \Sigma_2(\omega),
~~\frac{m^*(\omega)}{m_b} \approx 1+\lambda(\omega) \approx
1-\frac{\partial\Sigma_1(\omega)}{\partial\omega} .
 \ee
In the weak-coupling limit, where Eq.\ (\ref{EDMcondition}) is
well satisfied, the optical scattering rate $1/\tau(\omega)$ can
be approximately written as \cite{shulga2,allen,puchkov}
\be
 \label{coth}
\frac{1}{\tau(\omega)} &\approx& \frac{2\pi}{\omega}
\int_{-\infty}^{\infty}d\Omega \left [\omega \, \coth\left
(\frac{\Omega}{2T} \right) \right. \nonumber \\
&& \left. +(\omega-\Omega)\coth \left(\frac{\omega-\Omega}{2T}\right
)\right ] \alpha^2_{tr}F(\Omega) + \frac{1}{\tau_{imp}}\/,
\label{tau}
\ee
where $1/\tau_{imp}$ is the impurity contribution and
$\alpha^2_{tr}F(\omega)$ is a phonon density of states
weighted by the amplitude for large-angle scattering on the Fermi
surface, which has the same spectral structure as
$\alpha^2F(\omega)$, but their amplitudes can be lower. In this
paper we will not distinguish the $\alpha^2F(\omega)$ and
$\alpha^2_{tr}F(\omega)$ from now on. We note that there were
several previous attempts to invert $\alpha^2F(\omega)$
\cite{marsiglio,shulga}. For instance, one may obtain
$\alpha^2F(\omega)$ in the $T = 0$ limit of Eq.~(\ref{tau}) using
 \be
 \label{a2f-approx}
\alpha^2F(\omega) \approx \frac{1}{2\pi}\frac{d^2}{d\omega^2}
\left[ \frac{\omega}{\tau(\omega)} \right] \approx \frac{1}{2\pi}
\frac{\omega_p^2}{4\pi} \frac{d^2}{d\omega^2} Re \left[
\frac{\omega}{\sigma(\omega)} \right],
 \ee
where the second expression follows by using the EDM of Eq.\
(\ref{drude}). This formula, however, has limitations to be
applied to the experimental IR data because large error bars are
inevitable from the double differentials and it is valid only
when EDM is valid and at $T=0$ \cite{shulga}.


For strong-coupling FL where $\lambda(\omega)$ has a significant
$\omega$-dependence, analysis based on the EDM becomes less
accurate. For the marginal Fermi liquid, where $\Sigma_1(\omega)
\sim \omega\ln|\omega|$ and $\Sigma_2(\omega) \sim -|\omega|$,
the EDM is expected to give somewhat less reliable description
because $\Sigma_1 (\omega)$, unlike FL, deviates from the
linearity in $\omega$. The situation becomes worse for non-Fermi
liquid, where $\Sigma_1 (\omega) \sim -\omega^{1-\alpha}$ and
$\Sigma_2 (\omega) \sim -\omega^{1-\alpha}$ ($0<\alpha<1$), and
the EDM may give misleading and qualitatively incorrect
descriptions as were observed in the ruthnates
\cite{kostic,yslee}.

\section{Formulation of SELF-ENERGY ANALYSIS}

We now present how one can analyze the frequency-dependent
infrared conductivity with the formula of Eq.\
(\ref{calculation}). As is explained below, the problem is reduced
to a global minimization of a $N$-variable function. Using Eq.\
(\ref{calculation}), it is simple to calculate the conductivity
$\sigma(\omega)$ from a given self-energy $\Sigma(\omega)$. What
we are trying to do here is exactly the inverse of that: We wish
to extract $\Sigma(\omega)$ from an experimentally measured
$\sigma(\omega)$.

The SEA is implemented by defining the functional $W[\Sigma_2]$ as
 \begin{equation}
W[\Sigma_2] \equiv \int_0^{\omega_c} d\omega \left [\sigma_1
(\omega) - \sigma_1^{exp}(\omega) \right ]^2 \/,
 \label{master}
 \end{equation}
where $\sigma_1^{exp}$ and $\sigma_1 $ are the real parts of the
experimental data and calculated conductivity using Eq.\
(\ref{calculation}), respectively. Since the real part of
self-energy $\Sigma_1(\omega)$ can be obtained from the imaginary
part $\Sigma_2 (\omega) $ by the Kramers-Kronig transformation, we
can consider $\Sigma_2(\omega)$ as the only independent function.
The functional $W$ is positive definite, and has the global
minimum of zero for the self-energy which reproduces the
experimental conductivity data. The self-energies defined at the
$N$ discrete frequencies, $x_i \equiv \Sigma_2 (\omega_i)$ for
$i=1,\cdots,N$ ($\omega_N = \omega_c$), are taken as the $N$
independent variables of the function $W(x_1,x_2,\cdots,x_N)
\equiv W[\Sigma_2]$. The cutoff frequency $\omega_c$ is not
necessarily equal to the $\omega_{ max}$ of Eq.\
(\ref{sum-rule}). Now, the problem is reduced to a global
minimization of a $N$-variable function.

%
%

We note that, depending on the problems, other forms of the
functional $W$ may yield better results. For good metals, for
instance, whose mid-IR conductivity is very small compared with
far-IR region,
 \be
W=\int_0^{\omega_c} d\omega \, [\ln\sigma_1 (\omega)
-\ln\sigma_1^{exp}(\omega)]^2
 \ee
works better than the form of Eq.\ (\ref{master}). For the region
where $\omega>\omega_c$, which we need to know for the
Kramers-Kronig transformation, $\Sigma_2 (\omega)$ is taken as
constant. A constant $\Sigma_2 (\omega) $ for $\omega
>\omega_c$ corresponds to the reflectivity $R(\omega) \sim
1/\omega^4$, which is consistent with the standard procedure in
the IR experiments.

The global minimization of $W$ is achieved via the functional
derivative and numerical iterations. We start with an initial
configuration of $x_i^{(0)}$. A good initial guess can be
$x_i^{(0)}=-\frac{1}{2\tau(\omega_i)}$ obtained from EDM
analysis, or a negative constant value for the whole frequency
range. Then, we move to new $x_i$ along the steepest descent
direction of $W$.
 \be
\label{iteration} x_i^{(new)} = x_i^{(old)} - \left. s\ \frac{dW
}{dx_i}\right|_{x_i =x_i^{(old)}} ,
 \ee
where $\frac{dW }{dx_i}$ represents the functional derivative of
the functional $W[\Sigma_2]$ with respect to $\Sigma_2 (\omega)$
given by
 \be
\frac{d W}{dx_i}=2\int_0^{\omega_c} d\omega' \left[ \sigma_1
(\omega') -\sigma_1^{exp} (\omega') \right] \left. \Delta
\right|_{\omega=\omega_i},
\nonumber \\
\Delta \equiv
\frac{\delta\sigma_1(\omega')}{\delta\Sigma_2(\omega)}
+\frac{1}{\pi} \int_{-\infty}^{\infty} d\epsilon \frac{\delta
\sigma_1(\omega')}{\delta\Sigma_1(\epsilon)} {\cal P}
\frac{1}{\omega-\epsilon},
 \ee
where ${\cal P}$ stands for the principal value, and the use is
made of $\frac{\delta\Sigma_1(\epsilon)}{\delta\Sigma_2(\omega)}=
\frac{1}{\pi} {\cal P} \frac{1}{\omega-\epsilon}$. The step size
$s$ is chosen such that the $W$ is maximally decreased along the
steepest descent direction. The $\omega_p^2$, which sets the
scale of $\sigma(\omega)$ of Eq.\ (\ref{calculation}) is updated
at each iteration such that the calculated spectral sum is equal
to the experimental spectral sum up to $\omega_c$. Note that this
way of determining $\omega_p$ does $not$ require that $\omega_c =
\omega_{max}$ of Eq.\ (\ref{sum-rule}) and, therefore, the
extracted $\omega_p$ is almost independent of $\omega_{max}$ and
more reliable than that from the sum rule of Eq.\
(\ref{sum-rule}). This enables us to determine $\omega_p$ more
systematically as we will discuss in more detail below.

In general, the global minimization of a few hundred independent
variables is an extremely demanding problem. In the present case,
however, it is rendered tractable because of the following
observations: (1) We know the value of the global minimum unlike
general global minimization problems. It is exactly zero. (2) We
have some ideas about the physically meaningful form of the
$\Sigma_2 (\omega)$. It should be a continuous function of the
frequency and negative definite. (3) We have a better way to
escape from local minima than trying a new random starting point,
$x_i^{(0)}$: Take
$\Sigma_2^{(new)}(\omega)-\Sigma_2^{(old)}(\omega) \propto
\sigma_1(\omega)-\sigma_1^{exp}(\omega)$. An estimation of this
process can be obtained in following way. The contribution to the
conductivity at a given frequency $\omega$ is dominated by the
self-energy below $\omega$ in the low temperature limit due to the
thermal factor of $[f(\epsilon -\omega)-f(\epsilon)]/\omega$ of
Eq.~(\ref{calculation}) as discussed in Eq.\ (\ref{sigmaT0}). By
modifying the self-energy to be proportional to the difference
between calculated conductivity and the experimental data at each
frequency, we can continue our minimization modifying the `wrong'
region without spoiling a `good' region.

From the extracted $\Sigma(\omega)$, other physical properties
such as the plasma frequency, effective interaction between
electrons, Raman spectra, and inelastic neutron scattering
intensity can also be calculated. For instance, we use Eq.\
(\ref{a2F}) to find the effective interaction spectrum $P_2
(\omega)$ from an extracted $\Sigma_2(\omega)$ which was
discussed in the previous section. An important byproduct of the
SEA is that the plasma frequency $\omega_p$ can be determined
more accurately than has been $hitherto$ practiced, which is
necessary to determine the Drude parameters of Eq.\
(\ref{EDMpara}). Conventionally, $\omega_p$ is determined by
$\omega_{max}$ of Eq.\ (\ref{sum-rule}) which, in turn, is taken
such that the sum as a function of $\omega_{max}$ has the
smallest slope. This procedure can be problematic especially for
the materials without a sharp plasma edge in the reflectivity
data. In contrast, SEA works in this case as well because the
determination of $\omega_p$, which sets the scale of
$\sigma(\omega)$, is almost $independent$ of the $\omega_c$ of
the functional $W$. For instance, one can determine $\omega_p$
with only very restricted data between $0<\omega<\omega_c ~(\ll
\omega_{max})$. This will be illustrated for Nb below.

Another byproduct closely related to the $\omega_p$ determination
is that we may separate the intra- and inter-band contributions
more systematically. We can extract the self-energy of effective
single-component carriers by fitting the low-frequency
experimental conductivity up to $\omega_c \approx \omega_{max}$.
The conductivity $\sigma(\omega)$ calculated by substituting the
extracted $\Sigma(\omega)$ into Eq.\ (\ref{calculation}) is
intra-band conductivity. This process, however, is more sensitive
to $\omega_c$ than the $\omega_p$ determination.

We will demonstrate in the next section that the SEA is
straightforward to implement to extract the self-energy from
experimental IR conductivity without the local minimum problem.
We have found that the SEA yields the same solution $\Sigma_2
(\omega)$ from almost any initial configuration. The obtained
solution $\Sigma_2 (\omega)$, therefore, seems unique.

\section{Applications to experimental data}

We will now apply the SEA developed in the previous section to
real materials, and demonstrate that it can be readily employed
to analyze the frequency-dependent IR experimental data. In the
first two parts of this section, we will analyze the simple
metals, Pb and Nb. From the SEA, we can extract the self-energy
of the materials, and from the extracted self-energy we can find
the $\alpha^2 F(\omega)$ using Eqs.\ (\ref{a2F}) and
(\ref{alpha2f}). It can be compared with the measured $\alpha^2
F(\omega)$ well established from the tunneling experiments
\cite{wolf}. But for such a good metal, it is a very demanding
task to measure the conductivity in far-IR region, since the
reflectivity is very close to 100 \% there. Thus there exist few
published conductivity data for Pb and Nb. For Pb, we take the
experimental $\alpha^2 F(\omega)$ obtained from the tunneling
experiment and calculate the self-energy from Eqs.\ (\ref{a2F})
and (\ref{alpha2f}), and then calculate the conductivity from Eq.
(\ref{calculation}). This is taken as the ``experimental''
conductivity $\sigma^{exp}(\omega)$ for Pb. After that, the real
experimental data of Nb is analyzed \cite{pronin}. The results
are compared with those obtained from the tunneling spectra. The
last part of this section is devoted to the SEA for the normal
state IR conductivity data of MgB$_2$ superconductor. We will
argue from the SEA results that the small electron-phonon
coupling constant $\lambda$ extracted previously from the
$T$-dependence of the resistivity $\rho(T)$ and EDM analysis of
$\sigma(\omega)$, which is too small to account for the
superconducting transition temperature $T_c$, is most likely due
to an underestimate of the plasma frequency $\omega_p$.

\subsection{Application to Pb: Generated data}

We take Pb as a test case of the proposed SEA because its
$\alpha^2 F(\omega)$ is well established from the tunneling
\cite{wolf}. The electron-phonon coupling constant $\lambda$ of
this metal is known to be $\approx 1.5$. The frequency-dependent
IR conductivity of Pb, however, is not available. Therefore, the
IR conductivity data was generated with Eqs.\ (\ref{calculation})
and (\ref{a2F}) using the tunneling $\alpha^2F(\omega)$ at 200
frequencies up to $\omega_c = 15$ meV with an equal spacing. We
took the self-energy due to impurity scattering $\Sigma_2^{imp} =
-1 $ meV and $T = 3$ K as representative values. The generated
conductivity in this way, shown in the left column of Fig.~1 with
open circle, is taken as the ``experimental'' data $\sigma_1^{exp}
(\omega)$. It was then fitted by the SEA as explained above to
extract the self-energy $\Sigma_2 (\omega_i)$ without, of course,
any information about the self-energy used to generate the
conductivity. The results are shown in the solid line in the left
column of Fig.~1. The extracted and experimental conductivities
are almost indistinguishable. Note that the width (HWHM) of
$\sigma_1 (\omega)$ is not given by $-2\Sigma_2 (\omega=0)$
because of the substantial frequency dependence of $\Sigma_2
(\omega)$.

In the right column of Fig.\ 1, we show the real and imaginary
parts of the extracted self-energy in the forms of
$-2\Sigma_2(\omega)$ and $1-\Sigma_1(\omega)/\omega$, along with
the generated experimental data. The extracted and experimental
data are shown, respectively, by the solid line and open circles.
They are practically indistinguishable. For comparison, we also
plotted the EDM analysis of the experimental conductivity data,
$1/\tau(\omega)$ and $m^\ast(\omega)/m_b$, in dotted lines, using
Eq. (\ref{EDMpara}). As discussed in Eq. (\ref{mapping}),
$1/\tau(\omega)$ and $m^\ast(\omega)/m_b$ correspond,
respectively, $-2\Sigma_2(\omega)$ and $1-\Sigma_1
(\omega)/\omega$. The right panel demonstrate, however, that they
can substantially deviate from each other even for the relatively
simple metals. The EDM analysis in general yield smoother
frequency dependences for $1/\tau(\omega)$ and
$m^\ast(\omega)/m_b$. This is expected because $1/\tau(\omega)$ is
an ``average'' of the $\Sigma_2 (\omega)$ as discussed previously
in Section II. The curve shown by the thin solid line in the
lower picture was calculated by using the approximate formula
Eq.~(\ref{coth}), which is in a good agreement with the
$1/\tau(\omega)$ from the EDM analysis \cite{shulga2}.

From the extracted self-energy, $\alpha^2F(\omega)$ can be
obtained from a derivative with respect to $\omega$, $\alpha^2
F(\omega) = -\frac1\pi \frac{\partial \Sigma_2 (\omega)} {\partial
\omega} $, which is valid at low $T$ (dotted curve) or a
convolution of Eq.\ (\ref{a2F}) (thin solid curve) \cite{shulga}
as shown in the upper left inset of Fig.~1. The experimental
$\alpha^2 F(\omega)$ is shown in the thick solid line. The
extracted $\alpha^2 F(\omega)$ from the convolution is again
almost indistinguishable from the experimental one.
These results from the SEA show substantial improvements over the
previous attempts \cite{marsiglio} based on the assumptions of
both $T=0$ and weak-coupling limit. We argue from this example
that even for the simple metals, for which the EDM can give the
qualitatively valid description, the SEA can provide more
accurate and reliable description of the material, which can be
quite different from the EDM results.


\subsection{Application to Nb}

We now proceed to apply the present method to the experimental IR
data of Nb. We used the far-IR conductivity data measured by
Pronin {\it et al.} in the normal state at 9 K \cite{pronin}.
Their far-IR data are available from 84 up to 300 cm$^{-1}$ (with
constant data in the DC limit measured via a different method),
which is a good example to demonstrate that the $\omega_p$ can be
obtained without a full intra-band spectrum in the SEA.

We took $\omega_c = 250$ cm$^{-1}$ and $N=200$. For $0 \leq \omega
\leq $ 84 meV, we constrain $\Sigma_2 (\omega)$ to remain
constant. This implies vanishing $\alpha^2 F(\omega)$ in the
region as shown in the upper right inset of Fig.\ 2 and,
consequently, a reduced $\lambda$, and the Drude behavior in
$\sigma_1 (\omega)$. The fitted (solid curve) and experimental
(crosses) $\sigma_1 (\omega)$ are plotted in Fig.\ 2. The
electron-phonon coupling constant $\lambda =
-\partial\Sigma_1(\omega)/\partial\omega |_{\omega \rightarrow
0}$ from the SEA is 0.51 which is somewhat reduced compared with
the experimental value of $0.9-1$ \cite{wolf} as discussed above.
The calculated plasma frequency is 7.8 eV while the experimental
value is $7.2$ eV \cite{pronin}. In the lower left inset, the
extracted $\Sigma_2 (\omega)$ is shown. The above results clearly
demonstrate that the SEA works also for the case with only
restricted data.

These two examples of the SEA establish that the method can
indeed be applied to analyze the frequency conductivity and it
can provide the most microscopic information of interacting
electron systems, the self-energy $\Sigma(\omega)$ and the
effective interaction $P(\omega)$. The SEA can provide for the
Fermi liquids more reliable and accurate information than the
conventional EDM analysis. For the non-Fermi liquids, it is
expected to provide qualitatively different information which is
$not$ accessible with the EDM.

\subsection{Application to MgB$_2$}

Let us now analyze the IR data of normal state $c$-axis oriented
MgB$_2$ film measured by Tu {\it et al.} \cite{tu}, which we
regard as being due to intra-band excitations of an effective
single band system \cite{shulga}. In Fig.\ 3(a), the experimental
conductivities of Tu {\it et al.} at $T=45$ K and 295 K are shown
together with the fitted conductivities using the SEA with $N=300$
and $\omega_c = 6000$ cm$^{-1}$. The solid curves represent the
fitted conductivities, and the dashed and dotted curves,
respectively, the experimental ones at $T=295$ K and $T=45$ K.
The small discrepancies are due to phonons. In Fig.\ 3(b), the
results from the EDM analysis are shown in the left column, and
those from the present SEA in the right column. The solid and
dashed curves are, respectively, for $T=45$ K and 295 K. Tu {\it
et al.} found from the sum rule of Eq.\ (\ref{sum-rule}) that
$\omega_p$ is 14750 cm$^{-1}$ which yields, through the EDM, a
very weak electron-phonon coupling as shown in the left column of
Fig.\ 3(b). They obtained $\lambda_{tr} \approx 0.13$ using the
Bloch-Gr$\ddot{\rm u}$neisen formula to analyze the
$T$-dependence of the resistivity, which is consistent with the
EDM analysis \cite{tu,marsiglio2}. To the extent that
$\lambda_{tr} \approx \lambda$, it seems too small to account for
the superconducting transition temperature $T_c = 39$ K. On the
other hand, we found from the SEA that $\lambda \approx 0.56$ and
0.41 at, respectively, $T=45$ and 295 K as shown in Fig.\ 3(b),
which are substantially larger than what Tu {\it et al.} found.
The local density approximation calculation yields $\lambda_{tr}
\approx 0.6$ \cite{liu}. One way of seeing the discrepancy
between EDM and SEA is that the $\omega_p$ from SEA is larger
than that from EDM. We found that $\omega_p = 16690 ~{\rm
cm}^{-1}$ at $T=45$ K, and 16740 cm$^{-1}$ at $T=295$ K. If the
enhanced $\omega_p$ are used in the EDM, the resulting
$\lambda_{tr}$ are in good agreement with the SEA. As far as the
EDM and the $T$-dependence of the resistivity are concerned, the
enhanced $\omega_p$ resolves the problem of the small
$\lambda_{tr}$.

With the SEA, we may go further and perform the spectral analysis
to see the frequency range that contributes to $\lambda$, which
can $not$ be carried out with the EDM analysis. $\lambda =
2\int_0^{\infty} d\Omega \frac{\alpha^2 F(\Omega)}{\Omega}$,
where $\alpha^2 F (\omega)= \frac{1}{\pi}P_2 (\omega)$ may be
obtained from the extracted $\Sigma_2 (\omega)$ using Eq.\
(\ref{a2F}). The $-2\Sigma_2 (\omega)$ is shown in the right
column of Fig.\ 3(b). The extracted $\alpha^2F(\omega)$ is shown
in Fig.\ 4 together with those from tunneling \cite{tunneling}
(dashed line) and LDA calculation \cite{liu} (thin solid line).
They are shown in a wider frequency range in the inset. The
extracted $\alpha^2F(\omega)$ from SEA are characterized by two
frequency regions which make dominant contributions to $\lambda$;
around $\omega \approx$ 70 and 300 meV, which is, interestingly,
consistent with the model proposed by Marsiglio
\cite{marsiglio2}. The low frequency region around 70 meV is the
contribution from the $E_{2g}$ phonon mode, while the nature of
the high frequency region is not clear. The two regions contribute
almost equally to $\lambda $: $\lambda_{phonon} \approx 0.36
~(0.21)$ and the total $\lambda $ is 0.56 (0.41) for $T=45$ K
($T=295$ K).

The extracted $\alpha^2 F (\omega)$, compared with the local
density approximation (LDA) calculation, correctly captured the
main $E_{2g}$ contribution but missed smaller contributions from
other phonon modes; among the modes in the $\alpha^2 F (\omega)$
obtained from the LDA calculation shown in Fig.\ 4, the modes
whose $\alpha^2 F(\omega)$ are smaller than $\sim 0.5$ are absent
in the SEA results. This, we suspect, may be a consequence of the
broad features around 160 and 880 cm$^{-1}$ in the experimental
$\sigma(\omega)$, which were {\it not} predicted by phonon
calculations \cite{tu} and possibly due to MgO impurities. We did
not eliminate these contributions for the present calculations.
This may smear out otherwise sharper frequency dependence of
$\sigma(\omega)$, and reduce the electron-phonon coupling
constant.

%
%
Apart from these discrepances, the SEA successively describes the
frequency-dependent conductivity of the MgB$_2$ and yields an
increased $\lambda$ than the previous estimate. The SEA suggests,
which made more accurate determination of $\omega_p$ possible,
that the small $\lambda $ reported in the $c$-axis oriented
sample is most likely due to underestimated $\omega_p$, and the
total $\lambda $ from the SEA is substantially larger. However,
the $\lambda_{phonon} \approx \lambda /2$ still seems a bit too
small to account for the $T_c$.

\section{SUMMARY and discussions}

In this paper we have proposed and demonstrated a microscopic way
to analyze the frequency-dependent infrared conductivity,
referred to as self-energy analysis, which is valid for the
non-Fermim liquids as well as the Fermi liquids. Additional
advantage of the self-energy analysis is that the plasma
frequency $\omega_p$ can be obtained with a better accuracy.
Through the self-energy analysis, we extracted the electron
self-energy from the inversion of experimentally measured infrared
conductivity, and the effective interaction between electrons
from the extracted self-energy. After we demonstrated that the
self-energy analysis method can be successfully applied for
simple metals like Pb and Nb, we applied the method to the IR
conductivity data of normal state MgB$_2$. We have found that the
$\lambda $ from the self-energy analysis is substantially larger
than that obtained from the conventional analysis of the
$T$-dependence of the resistivity and extended Drude model. The
discrepancies between the self-energy analysis and conventional
method for MgB$_2$ is attributed to the underestimated $\omega_p$
from the conventional method. However, the $\lambda_{phonon} $
from the self-energy analysis still seems a bit small to account
for the $T_c$.

Now that we have demonstrated that the self-energy analysis of
the frequency-dependent infrared conductivity really works and can
be very powerful, some concluding remarks and outlooks are in
order. First, one may wonder if the solution to the global
minimization of the functional $W[\Sigma_2]$ is unique. The
answer seems positive: the converged solutions, with different
initial configurations, of given conductivity data all agree with
each other. This means that the extracted self-energy $\Sigma_2
(\omega)$ is indeed unique. Second, we have also checked if there
is any spurious feature from the Kramers-Kronig transformation
used in the present work to obtain the real part of the
self-energy from the imaginary part. We therefore have fitted the
complex conductivity by treating both the real and imaginary parts
of $\Sigma(\omega)$ as independent variables, which eliminates the
use of Kramers-Kronig. This gives the same $\Sigma_2 (\omega)$
with the procedure using $W[\Sigma_2]$, demonstrating the
reliability of the present method. Third, we plan to report
results of the self-energy analysis of the conductivity for other
correlated electron systems such as the high $T_c$
superconductors, Sm$_2$/Nd$_2$Mo$_2$O$_7$, and Ca/SrRuO$_3$
mentioned in the introduction. The differences between the
self-energy analysis and the extended Drude model analysis are
expected to be $quantitative$ for the Fermi liquids, but be
$qualitative$ for the non-Fermi liquids. It will therefore be
very interesting to see what information the self-energy analysis
provide for the strongly correlated electron systems. Also among
the plan are extending the self-energy analysis to non constant
density of states and broken symmetry states, and doing the
analysis for other two-particle probes such as the electronic
Raman and inelastic neutron spectra. We believe that the
self-energy analysis seems timely and urgent in view of the
mounting interests in the correlated electron or non-Fermi liquid
systems for which the phenomenological analyses may yield
inadequate results.

\acknowledgments

We would like to thank J.~J.~Tu and A.~V.~Pronin for providing
their experimental data for our analyses. We also thank Tae Won
Noh, Jae-Hoon Kim, In-Sang Yang, Jaejun Yu, Yunsang Lee, Mi-Ock
Mun, Jooyoung Lee, Seki Kim, M. J. Rice, D. van der Marel, D. B.
Tanner, T. Timusk, F. Marsiglio, S. V. Shulga, and S. L. Cooper
for helpful comments and discussions. This work was supported by
the Korea Science \& Engineering Foundation (KOSEF) through grant
No.\ R01-1999-000-00031-0 and CNNC, and by the Ministry of
Education through BK21 SNU-SKKU program.

\begin{figure}
\label{fig1}
\vspace{-0.8cm}
\includegraphics[scale=0.9]{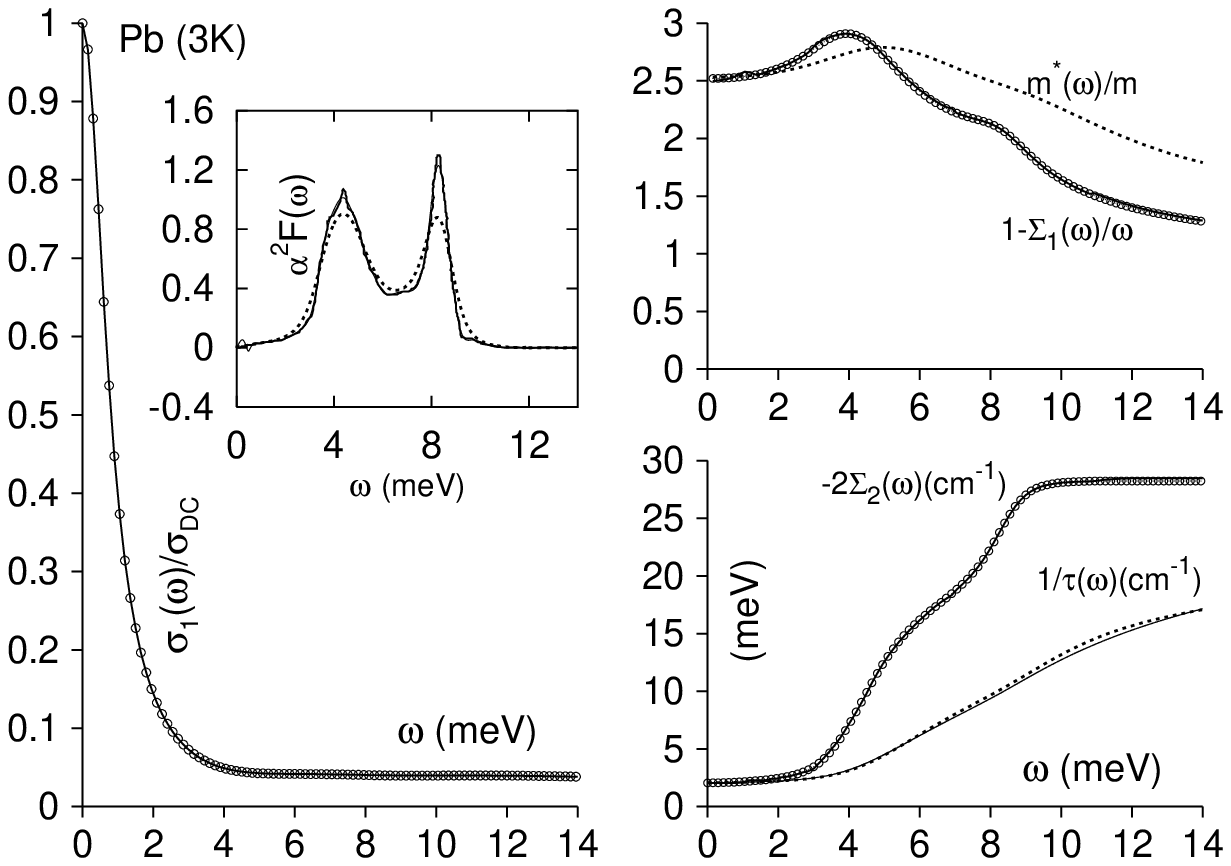}
\vspace{1.5cm}
\caption{Left: $\sigma_1 (\omega)$ is plotted for Pb with
$\Sigma_2^{imp} = -1 $ meV and at $T=3$ K. The extracted and
generated conductivities are shown, respectively, in solid line
and open circles. The dashed and thin solid curves in the inset
were obtained from, respectively, $\alpha^2 F(\omega) =
-\frac{1}{\pi} \frac{\partial \Sigma_2 (\omega)}{\partial\omega}$
and the convolution using Eq.\ (\ref{a2F}), and the latter is
almost indistinguishable with the input tunneling data (solid
line). Right: The $1-\Sigma_1(\omega)/\omega$ and $-2\Sigma_2
(\omega)$ obtained from the SEA are plotted, respectively, in the
upper and lower panels. The extracted (generated) ones are shown
in solid line (open circles). These SEA analysis results are
compared with the EDM results. The dotted lines represents
$1/\tau(\omega)$ and $m^\ast(\omega)/m_b$, respectively, obtained
from EDM using the generated conductivity. $1/\tau(\omega)$ (thin
solid line) calculated by using Eq.\ (\ref{coth}) is also shown
in the lower right panel for comparison. }

\end{figure}

\begin{figure}
\label{fig2}
\includegraphics[scale=0.9]{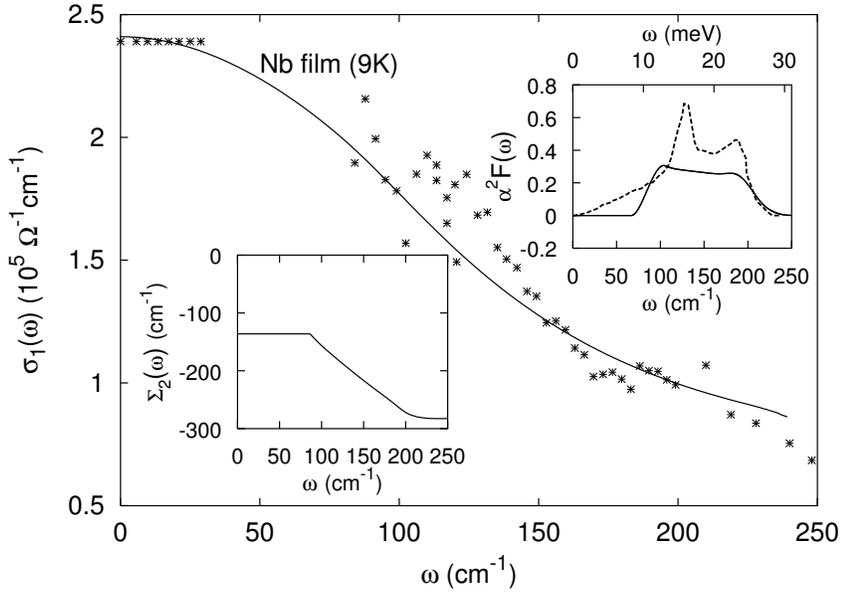}
\vspace{1.5cm}
 \caption{The calculated and experimental $\sigma_1
(\omega)$ of Nb are shown with solid curve and crosses,
respectively. In the insets, the extracted $\alpha^2F(\omega)$
and $\Sigma_2 (\omega)$ are shown along with the tunneling
$\alpha^2F(\omega)$ (dashed).}
\end{figure}

\begin{figure}
\label{fig3}
\includegraphics[scale=0.9]{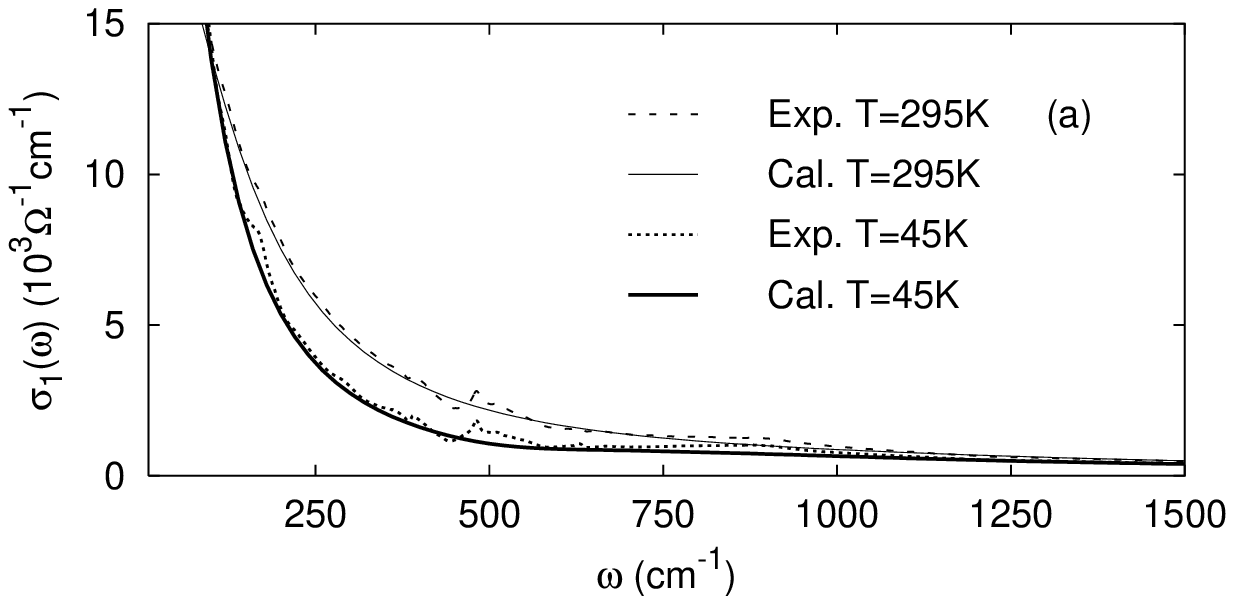}
\includegraphics[scale=0.9]{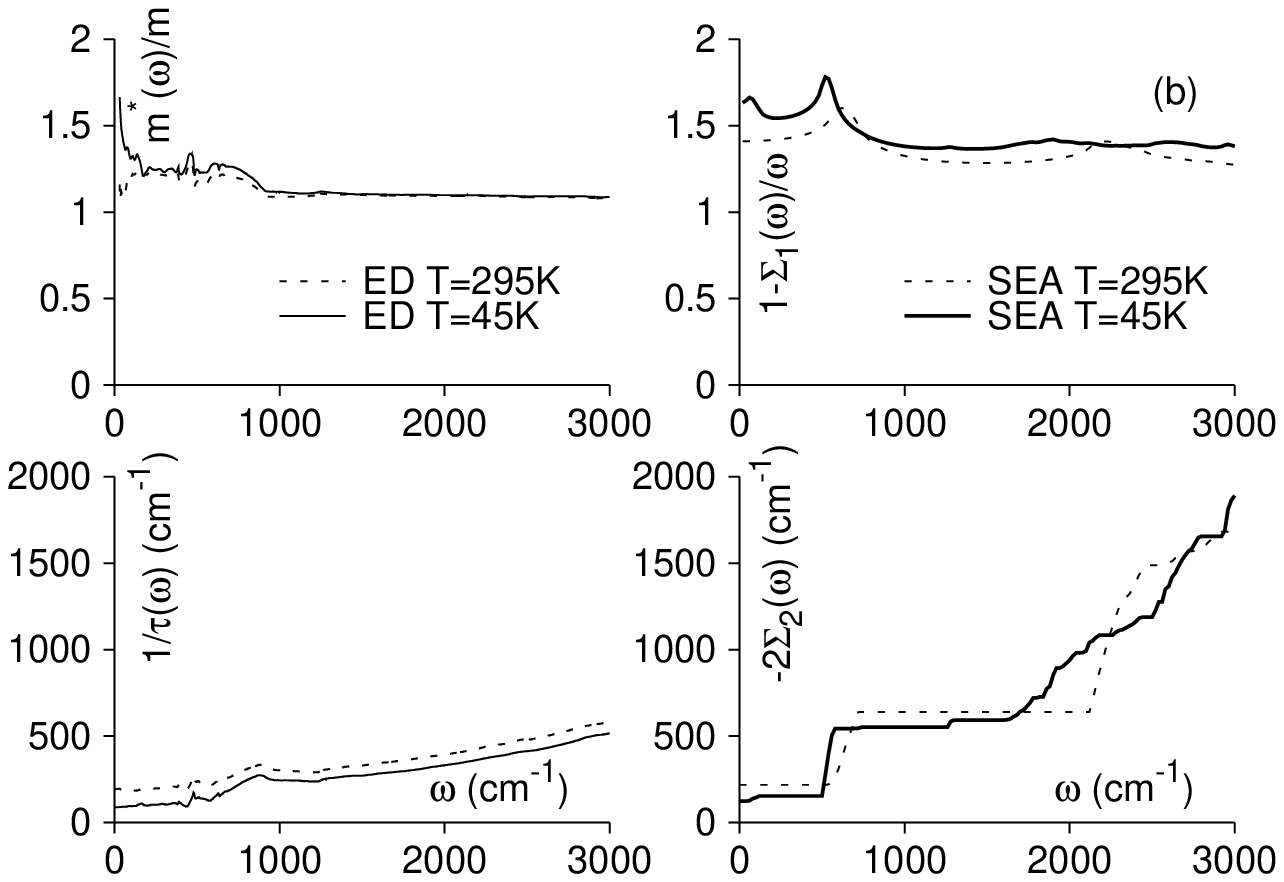}
\vspace{1.5cm}
 \caption{(a) The calculated and measured $\sigma_1
(\omega)$ of MgB$_2$ for $T = 45$ and $295$ K. The small
discrepancies are due to phonons. (b) Left: extended Drude
analysis of experimental data indicating the weak electron-phonon
coupling. Right: the corresponding quantities from the self-energy
analysis. Note that $m^*/m_b$ and $1/\tau(\omega)$ show much
smoother frequency dependence compared with the corresponding
$1-\Sigma_1 (\omega)$ and $-2\Sigma_2 (\omega)$ as noted
previously for Pb in Fig.\ 1.}
\end{figure}

\begin{figure}
\label{fig4}
\includegraphics[scale=0.9]{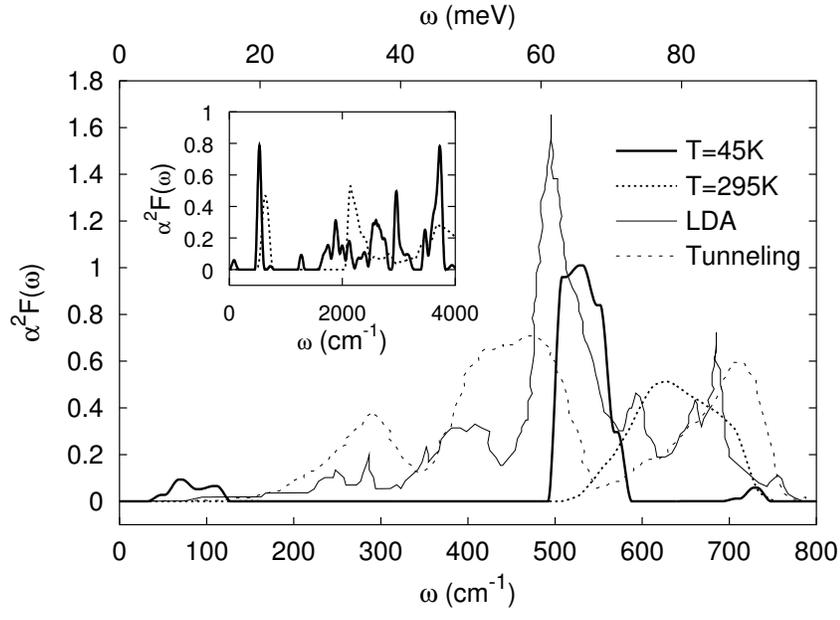}
\vspace{1.5cm}
 \caption{The extracted $\alpha^2F(\omega)$ from $\Sigma_2 (\omega)$ of
Fig.\ 3 along with those from LDA and tunneling. The extracted
$\alpha^2F(\omega)$ shown in a wider frequency range in the inset
are characterized by two dominant contributions around $\omega
\approx $ 70 and 300 meV.}
\end{figure}

\end{document}